\pdfoutput=1
\documentclass{elsarticle}
\usepackage{hyperref}
\journal{Journal of Astroparticle Physics}

\usepackage{upgreek} 
\usepackage{booktabs}

\newcommand{\ttwo}{\mbox{T$_2$}}

\newcommand{\mnu}{m(\overline{\upnu}_e)}

\graphicspath{{figures/}}

\begin{document}

\begin{frontmatter}

\title{Deconvolution of the energy loss function of the KATRIN experiment}

\author{V.~Hannen} \ead{hannen@uni-muenster.de}
\author{I. Heese}
\author{C.~Weinheimer}
\address{Institut f\"ur Kernphysik, Westf\"alische Wilhelms-Universit\"at M\"unster, Wilhelm-Klemm-Str.~9, 48149 M\"unster, Germany}

\author{A.~Sejersen~Riis}
\address{Department of Physics and Astronomy, University of Aarhus, Ny Munkegade, DK-8000 Aarhus C, Denmark}

\author{K.~Valerius}
\address{Institut f\"ur Kernphysik, Karlsruher Institut f\"ur Technologie, Hermann-von-Helmholtz-Platz~1, 76344 Eggenstein-Leopoldshafen, Germany}

\begin{abstract}
The KATRIN experiment aims at a direct and model independent determination of the neutrino mass with 0.2~eV/c$^2$ sensitivity (at 90\% C.L.) via a measurement of the endpoint region of the tritium beta-decay spectrum. The main components of the experiment are a windowless gaseous tritium source (WGTS), differential and cryogenic pumping sections and a tandem of a pre- and a main-spectrometer, applying the concept of magnetic adiabatic collimation with an electrostatic retardation potential to analyze the energy of beta decay electrons and to guide electrons passing the filter onto a segmented silicon PIN detector.\\
One of the important systematic uncertainties of such an experiment are due to energy losses of $\upbeta$-decay electrons by elastic and inelastic scattering off tritium molecules within the source volume which alter the shape of the measured spectrum. To correct for these effects an independent measurement of the corresponding energy loss function is required. In this work we describe a deconvolution method to extract the energy loss function from measurements of the response function of the experiment at different column densities of the WGTS using a monoenergetic electron source.
\end{abstract}


\end{frontmatter}

%\linenumbers

%
\section{Introduction}
The KArlsruhe TRItium Neutrino (KATRIN) experiment aims at determining the neutrino mass in a model independent way from the kinematics of tritium $\upbeta$-decay. The observable in this case is an ''average electron anti-neutrino mass'' given by the incoherent sum of neutrino mass eigenstates weighted by the squared elements of the mixing matrix. % (see equation~\ref{eq:numass}).
The experiment combines a Windowless Gaseous Tritium Source (WGTS) and a high resolution electrostatic retarding spectrometer (MAC-E filter) to measure the spectral shape of $\upbeta$-decay electrons close to the endpoint energy at 18.6~keV with an unprecedented precision. KATRIN's sensitivity to the neutrino mass will be 0.2~eV/c$^2$ (at 90\% C.L.) after 3 years worth of data taking~\cite{KAT04}. 
An observed mass signal of 0.35~eV/c$^2$ will have a $5\sigma$ significance at the expected level of statistic and systematic uncertainties. 
In order to reach the desired sensitivity, all systematic effects of the measurement must be well under control with the major systematic uncertainties being allowed to contribute no more than $\Delta m^2 = 0.0075~{\rm eV}^2/{\rm c}^4$ to the systematic error budget.
\par
\begin{figure}[h]
\centering
\includegraphics[width=0.95\textwidth]{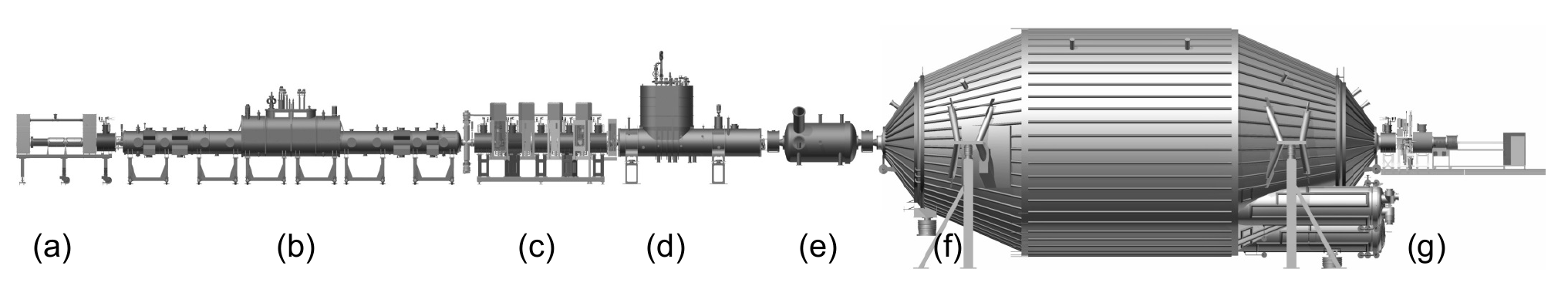}
\caption{Overview of the KATRIN experiment. The main components are: (a) calibration and monitoring system, (b) windowless gaseous tritium source, (c) differential and (d) cryogenic pumping sections, (e) pre-spectrometer, (f) main spectrometer, (g) detector system. Overall length ca. 70~m.
\label{fig:katrin}}
\end{figure}

An overview of the KATRIN experiment is shown in figure~\ref{fig:katrin}. The experiment starts with the WGTS where 
molecular \ttwo\ gas is injected at the center of the source and removed at both ends by turbo-molecular pumps. The \ttwo\ gas is kept at a constant temperature of 30~K within the source that is operated at a column density of $5\cdot 10^{17} {\rm cm}^{-2}$. The operational parameters of the source cryostat are monitored by a complex sensor network and a dedicated calibration and monitoring section at the rear of the source system~\cite{WGTS}. 
About $10^{10}$ $\upbeta$-decay electrons are emitted per second into the accepted forward solid angle with pitch angles less than $\theta_\mathrm{max} = 51^\circ$ and are guided magnetically through the transport section to the spectrometer tandem consisting of pre- and main-spectrometers. The task of the transport section made up of a differential pumping section and a cryo-pumping section is to suppress the flow of \ttwo\ molecules into the direction of the spectrometers by at least a factor of $10^{14}$ in order to reduce experimental background from tritium decays within the spectrometers. A first energy discrimination is performed by the pre-spectrometer which rejects the low energy part of the $\upbeta$ spectrum (up to 300~eV below the endpoint) and thereby reduces the rate of electrons going into the main spectrometer to approximately~$10^3\,{\rm s}^{-1}$. Like the pre-spectrometer the main spectrometer operates as a so-called MAC-E filter~\cite{Pic92} and has the task to perform a precise energy analysis of the decay electrons.
\par
In a MAC-E filter electrons are guided magnetically against an electrostatic retardation potential that can only be surpassed by electrons with  sufficiently high longitudinal energy with respect to the electric field. Here the longitudinal energy is given by $E_\parallel = E_{\rm kin} \cdot \cos^2{\theta}$ with $E_{\rm kin}$ being the kinetic energy of the electron and $\theta$ being the angle between electron momentum and  magnetic field direction. The transverse energy is accordingly given by $E_\perp =  E_{\rm kin} \cdot \sin^2{\theta}$.
The spectrometer acts as a high pass filter with a transmission function describing the observed electron rate as a function of the electron surplus energy (see section~\ref{sec:resp}). To reduce the amount of transversal energy of the electrons that is not analyzed by the spectrometer, the technique of magnetic adiabatic collimation is used. The idea is that the magnetic guiding field drops by several orders of magnitude from the entrance of the spectrometer to the analyzing plane, where the electric potential reaches its maximum. If the gradient of the magnetic field is small enough, such that the field is approximately constant along one cyclotron loop of the electron movement, the magnetic moment of the cyclotron motion $\mu = E_\perp / B$ (non-relativistic) is constant, and as $B$ drops the transversal energy of the electrons is converted into longitudinal energy $E_\parallel$ that can be analyzed by the spectrometer. 
By varying the electric potential of the spectrometer it is then possible to scan the relevant region around the endpoint energy of tritium $\upbeta$-decay and accumulate a spectrum. 
Electrons with sufficient energy to pass the spectrometer are finally detected by a 148 pixel silicon PIN detector~\cite{Ams15} at the end of the setup.
\par
Among the main systematical uncertainties of the experiment are energy losses from inelastic scattering of electrons in the source, fluctuations of the source density, fluctuations of the spectrometer analyzing potential, uncertainties in the transmission function and uncertainties in the final state distribution of the daughter molecules left after the decay reaction. A sophisticated calibration and monitoring system is being set up to keep the aforementioned systematic effects under control.
While there is some information on the energy loss of 18.6~keV electrons in gaseous tritium or quench condensed deuterium from the former neutrino mass experiments in Troitsk and Mainz~\cite{Ase00}, precise experimental information on energy losses of electrons with energies near the endpoint of the tritium $\upbeta$ spectrum are only available for molecular hydrogen as target gas~\cite{Gei64,Uls72}.
\par
A measurement of the energy differential scattering cross section of 18.6~keV electrons off molecular tritium is therefore highly desirable. Such a measurement can be performed using a monoenergetic source of electrons mounted upstream of the WGTS to determine the response function of the overall experiment at different column densities of the source. A deconvolution method suitable to extract the energy loss function from the data will be presented in the following sections.
\par
Once the energy loss function for tritium is known with sufficient accuracy, the same measurement setup can be used for an independent check of the column density of the WGTS during intervals between the regular measurement cycles of the KATRIN experiment~\cite{KAT04}. 
\section{Energy loss function}
\label{sec:eloss}
The processes contributing to the energy loss of electrons traversing the molecular tritium gas within the WGTS are excitation of rotational and vibrational states of the T$_2$ molecules, excitation of electronic molecular states, dissociation and ionization of the molecules.
\par
Aseev et al.~\cite{Ase00} report on measurements of energy losses of electrons in gaseous tritium and in quench condensed deuterium films. Because of the limited energy resolution of a few eV the shape of the energy loss spectrum was not directly extracted from the data in their analysis, but approximated by a Gaussian representing electronic excitations and dissociation and a one-sided Lorentzian curve representing the continuum caused 
by ionization of the molecules. The parameters of the two functions were then adapted to fit the observed integral energy spectra obtained with an 18.6~keV mono-energetic electron source for gaseous tritium or from 17.8~keV mono-energetic conversion electrons from a $^{83\rm m}$Kr film covered by various thicknesses of D$_2$ absorbers. In both cases, energy losses caused by rotational and vibrational excitations of the molecules without electronic excitation could not be resolved and were neglected.
\par
\begin{figure}[h]
\centering
\includegraphics[width=1.0\textwidth]{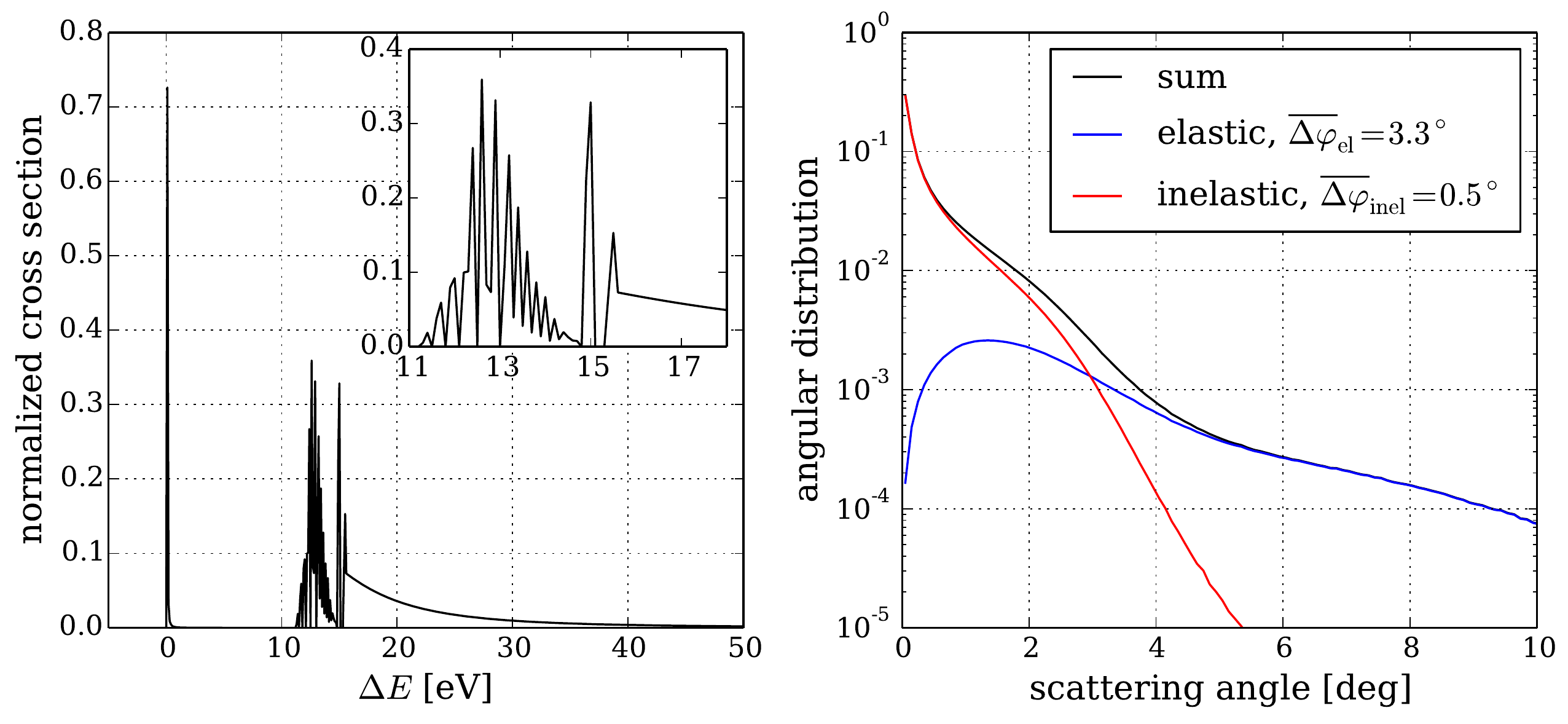}
\caption{Left: Normalized energy loss function $f(\Delta E)$ for scattering off hydrogen according to~\cite{Glueck} (bin width 0.1~eV). 
Right: corresponding angular distribution for elastic and inelastic scattering, normalized to 1.}
\label{fig:eloss_fcn}
\end{figure}

More detailed information is available for the scattering of 25~keV electrons from molecular hydrogen gas~\cite{Gei64,Uls72} where direct measurements of the energy loss function with resolutions down to 40~meV have been performed. % (see figure~\ref{fig:geiger}). 
%
%\begin{figure}[h]
%\centering
%\includegraphics[width=0.36\textwidth]{geiger-1.png}
%\includegraphics[width=0.62\textwidth]{geiger-2.png}
%\caption{Left: energy loss of 25~keV electrons in molecular hydrogen gas for small scattering angles at an experimental resolution of $\approx 1$~eV. Right: energy loss of 25~keV electrons near the threshold energy for electronic excitation of the hydrogen molecules measured with $\approx 0.04$~eV resolution (reprinted from~\cite{Gei64}, Copyright 1964, with permission of Springer).}
%\label{fig:geiger}
%\end{figure}
%
This information about the scattering of electrons from molecular hydrogen has been implemented into a computer code by F.~Gl\"uck~\cite{Glueck} that can be used in simulations to generate energy losses $\Delta E$ and scattering angles $\Delta \varphi$ in individual scattering events. The spectral shape produced with this routine is shown in figure~\ref{fig:eloss_fcn}. 
It is used in a toy Monte Carlo simulation of the WGTS to evaluate the deconvolution methods described in the following sections.
\par
The probability for an electron of kinetic energy $E$ to lose a specific amount of energy $\Delta E$ in a single scattering event is described by the differential energy loss function $\frac{\rm d\sigma}{\rm d\Delta E}$. For our purpose, we normalize the function by the total inelastic scattering cross section $\sigma_{\rm tot}$, obtaining\footnote{The integral over the energy losses runs up to $E/2$ since the incoming electron and the secondary electron in an ionisation process (assuming $E$ is larger than twice the ionisation energy) are identical quantum particles.}
\begin{equation}
  f(\Delta E) = \frac{1}{\sigma_{\rm tot}} \cdot \frac{\rm d\sigma}{\rm d\Delta E} \qquad {\rm with} \qquad
  \int^{E/2}_0 f(\Delta E) \, \rm d\Delta E = 1 \; .
\end{equation}
The total inelastic scattering cross section for 18.6~keV electrons off gaseous tritium is given by $\sigma_{\rm tot}({\rm T_2}) = (3.40 \pm 0.07)\cdot 10^{-18}~{\rm cm}^2$~\cite{Ase00}. 
The above mentioned code by F.~Gl\"uck~\cite{Glueck} for scattering of 18.6~keV electrons off hydrogen gives a total inelastic cross section of $\sigma_{\rm tot}({\rm H_2}) = 3.7 \cdot 10^{-18}~{\rm cm}^2$.
\section{Deconvolution method}
In the following sections we describe suitable mathematical methods to extract the energy loss function of 18.6~keV electrons in gaseous tritium from a series of measurements of the overall response function of the experiment at different column densities of the WGTS.
\subsection{Response function}
\label{sec:resp}

\begin{figure}[h]
 \centering
 \includegraphics[width=0.6\textwidth]{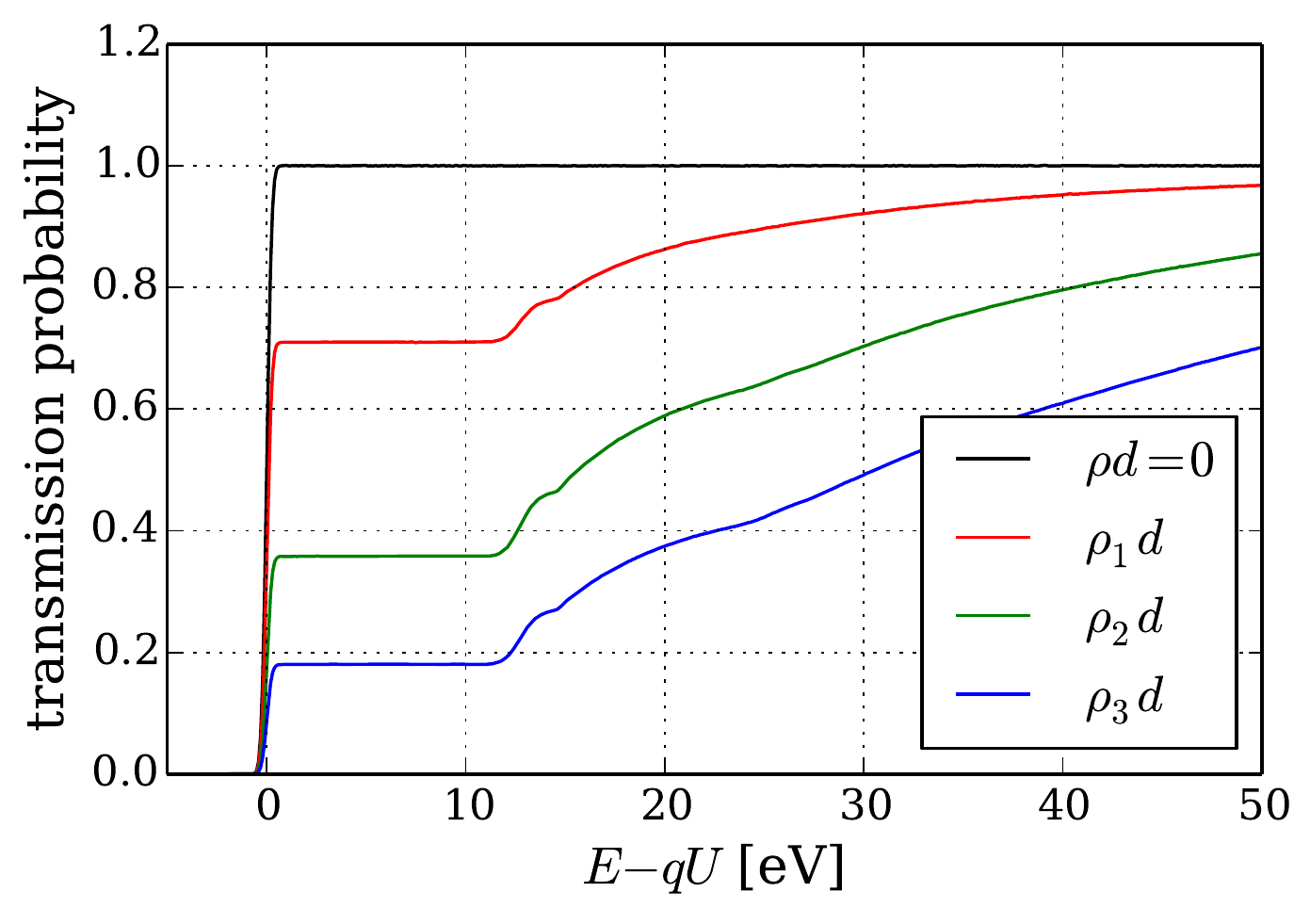}
 \caption{Simulated response functions of the KATRIN experiment for different column densities.}
 \label{fig:fres}
\end{figure}
Figure~\ref{fig:fres} displays simulated response functions of the KATRIN experiment at different column densities $\rho d = 0$ and %
$\rho_1 d < \rho_2 d < \rho_3 d$ assuming a mono-energetic electron source with narrow angular emission characteristics, i.e. pitch angles w.r.t. the magnetic field lines $\theta \leq {\cal O}(1^\circ)$. 
Shown is the transmission probability as a function of the nominal surplus energy $E_s = E - qU$ of the electrons, i.e., the difference between the setpoint energy $E$ of the electron source and the retardation potential $qU$ of the main spectrometer. 
\par
The response function at non-zero column density of the tritium source is given by a summation over contributions corresponding to n-fold (i.e. no, single, double, etc.) scattering of the electrons within the tritium source, weighted by the probabilities $P_n$ for n-fold scattering
\begin{equation}
  R(E_s) = P_0 \cdot T_e(E_s) + P_1 \cdot T_e(E_s) \otimes f(\Delta E) + 
            P_2 \cdot T_e(E_s) \otimes  f(\Delta E) \otimes f(\Delta E) + \ldots \; .
\label{eq:resp1}
\end{equation}
Here $T_e(E_s)$ is the experimental transmission function of the main spectrometer for the given electron source and $f(\Delta E)$ the sought after energy loss function neglecting small scattering angles. 
Given that we only consider a small energy interval of up to 50~eV below the endpoint energy of 18.6~keV of the $\upbeta$ spectrum, the dependence of the energy loss function $f(\Delta E)$ on the kinetic energy of the electrons can be neglected.
The scattering probabilities are normalized such that $\sum_{n=0}^\infty P_n = 1$.
The experimental transmission function $T_e(E_s)$ is determined from a measurement of the response function $R(E_s)$ with an empty tritium source and hence without scattering. It is then equal to the analytical transmission function of the spectrometer $T(E_s)$ convolved with a function $S_e$ describing the energy spread and angular distribution of the electron source
\begin{equation}
 T_e(E_s) = \left. R(E_s)\right|_{\rho d=0} = T(E_s) \otimes S_e \; .
\end{equation}
In the simulations, the energy spread of the source is described by a Gaussian smearing of the energy setpoint with a width of $\sigma_e = 0.2$~eV. It is assumed that the electron source has a small angular divergence with starting angles $\theta_e \leq 0.5^\circ := \theta_\mathrm{e,max}$ relative to the magnetic field direction at its location at the rear end of the WGTS. Within this narrow cone, the emission angles of the electrons are assumed to be isotropically distributed. 
Suitable electron sources that emit single electrons at adjustable total energy and adjustable emission angle have been developed and tested within the KATRIN collaboration~\cite{Val11,Bec14}.
The above mentioned numerical values for the energy and angular spread are compatible with the characteristics of the photo-electron source used during the commissioning of the KATRIN main spectrometer~\cite{Beh16}.
\par
The analytical transmission function of the spectrometer $T(E_s)$ for such a source is given by the following relation~\cite{Ase00}
\begin{equation}
 T(E_s) = \left\{
  \begin{array}{ll}
   0 & {\rm for}~~ E < qU \\
%   \left(1 - \sqrt{1 - \frac{E-qU}{E} \frac{B_e}{B_A}}\;\right) \Big/ \left(1 - \sqrt{1 - \frac{E_{\perp,A,{\rm max}}}{E} 
%     \frac{B_e}{B_A}}\;\right)~~ & {\rm for}~~ qU \leq E \leq qU + E_{\perp,A,{\rm max}}\\
   \frac{1 - \sqrt{1 - \frac{E-qU}{E} \frac{B_e}{B_A}}}{1 - \sqrt{1 - \frac{E_{\perp,A,{\rm max}}}{E} \frac{B_e}{B_A}}}~~ & {\rm for}~~ qU \leq E \leq qU + E_{\perp,A,{\rm max}}\\
   1 & {\rm for}~~ qU + E_{\perp,A,{\rm max}} < E,
  \end{array}
 \right. 
\end{equation}
where $B_e$ is the magnetic field at the electron source, $B_A$ the magnetic field at the analyzing plane of the spectrometer and $E_{\perp,A,{\rm max}} =  E\sin^2(\theta_{e,{\rm max}})\frac{B_A}{B_e}$ is the maximum remaining transversal energy component in the analyzing plane.\\
Defining
\begin{eqnarray}
 \epsilon_0(E_s) & = & T_e(E_s) \nonumber\\
 \epsilon_1(E_s) & = & T_e(E_s) \otimes f(\Delta E) \\
 \epsilon_2(E_s) & = & T_e(E_s) \otimes f(\Delta E) \otimes f(\Delta E) \nonumber\\
  & \ldots & \nonumber
\end{eqnarray}
as the n-fold scattering functions we can rewrite equation~\ref{eq:resp1} to obtain
\begin{equation}
 R(E_s)= P_0 \cdot \epsilon_0(E_s) +  P_1 \cdot \epsilon_1(E_s) +  P_2 \cdot \epsilon_2(E_s) + \ldots \; .
\end{equation}
If we manage to determine the single scattering function $\epsilon_1(E_s)$ from measured response functions we can, with the knowledge of $T_e(E_s)$, extract the energy loss function $f(\Delta E)$ using suitable deconvolution methods. 
\subsection{Scattering probabilities}
The mean free path of the electrons within the tritium gas inside the WGTS can be expressed in terms of a mean free column density $(\rho d)_{\rm free}$ which the electrons pass before an interaction and which is calculated taking the inverse of the total scattering cross section $(\rho d)_{\rm free} = 1/\sigma_{\rm tot}$.
The actual column density seen by an electron traversing the WGTS at an angle $\theta$ relative to the symmetry axis, i.e. the magnetic field axis, is given by $\rho d / \cos\theta$. Neglecting possible scattering angles $\Delta \varphi$ in the scattering processes for the moment, the mean number of expected scatterings is
\begin{equation}
 \mu(\theta) = \frac{\rho d}{(\rho d)_{\rm free} \cos\theta} =  \frac{\rho d \; \sigma_{\rm tot}}{\cos\theta} 
             = \frac{\mu_0}{\cos\theta}~ .
\end{equation}
The probability for an n-fold scattering is given by a Poissonian distribution:
\begin{equation}
 P_n(\mu(\theta)) = \frac{\mu^n(\theta)}{n!} \exp(-\mu(\theta)) \quad {\rm with} \quad n = 0,\; 1,\; 2,\; \ldots \; .
 \label{eq:Pn}
\end{equation}
We have to take into account that electrons generated by the electron source follow an angular distribution which is, for our purpose, assumed to be isotropic within a narrow interval between $0^\circ \leq \theta_e \leq \theta_{e,{\rm max}}$. 
If we further take into account that the magnetic field at the location of the electron source will be lower than the field within the WGTS, the starting angles have to be transformed according to 
\begin{equation}
 \theta = \arcsin \left( \sin \theta_e \cdot \sqrt{\frac{B_{\rm WGTS}}{B_e}} \right) \approx \theta_e \cdot \sqrt{\frac{B_{\rm WGTS}}{B_e}} \; ,
\end{equation}
resulting in angles $\theta$ of electron momenta relative to the magnetic field direction within the WGTS.\\
To obtain average scattering probabilities we weigh the values from equation~\ref{eq:Pn} with $g(\theta) = \sin \theta$, corresponding to the probability function of an isotropic distribution, integrate over the given range of angles and normalize 
\begin{eqnarray}
\label{eq:Pnav}
 \overline{P}_n(\mu_0) & = & \int^{\theta_{\rm max}}_0 g(\theta) P_n(\mu(\theta)) \, \rm d\theta ~ \Big/ \int^{\theta_{\rm max}}_0 g(\theta) \, \rm d\theta \\
 & = & \int^{\theta_{\rm max}}_0 \sin \theta \cdot \frac{(\mu_0/\cos\theta)^n}{n!} \exp(-\mu_0/\cos\theta) \, \rm d\theta \; \Big/ \;
       (1 - \cos \theta_{\rm max}) \nonumber
 \; .
\end{eqnarray}
Equation~\ref{eq:Pnav} delivers approximate values for the scattering probabilities, as it does not take into account changes in the direction of the electrons during scatterings. Secondly this equation also assumes a homogeneous distribution of tritium molecules in transverse direction within the WGTS. 
Compared to scattering probabilities extracted from the simulations accounting for scattering angles in elastic and inelastic scattering described in section~\ref{sec:MC} the deviations to the results calculated with~\ref{eq:Pnav} were found to be on the $<10^{-3}$ level, however. The small difference is due to the fact, that the scattering angles for inelastic scattering in our energy loss range of interest ($\Delta E < 50$~eV) are strongly forward peaked with a mean of $\overline{\Delta\varphi}=0.5^\circ$ and elastic scattering is a subdominant process (see figure~\ref{fig:eloss_fcn}, right). \\
To obtain more precise values for the scattering probabilities, a detailed simulation using a 3-dimensional description of the column density within the WGTS is required, which is beyond the scope of this paper.
\subsection{Extraction of the single scattering function}
\label{sec:scat}
In order to determine the single scattering function $\epsilon_1(E_s)$ we have to perform measurements of the response function $R(E_s)$ at different column densities. Neglecting multiple scattering events with more than three interactions of the electrons with the tritium gas inside the WGTS\footnote{The probability of a 4-fold scattering process with the maximum energy loss under consideration $\Delta E < 50$~eV is below 8\% at the maximum column density of $5\cdot 10^{17}~{\rm cm}^{-2}$ used in the simulations.}, we can set up a system of linear equations for measurements at three column densities labeled $a$, $b$ and $c$:
\begin{eqnarray}
\centering
  R^a(E_s) - P_0^a \cdot T_e(E_s) & = & P_1^a \cdot \epsilon_1(E_s) + P_2^a \cdot \epsilon_2(E_s) + P_3^a \cdot \epsilon_3(E_s) \nonumber \; ,\\
  R^b(E_s) - P_0^b \cdot T_e(E_s) & = & P_1^b \cdot \epsilon_1(E_s) + P_2^b \cdot \epsilon_2(E_s) + P_3^b \cdot \epsilon_3(E_s)  \nonumber \; ,\\
  R^c(E_s) - P_0^c \cdot T_e(E_s) & = & P_1^c \cdot \epsilon_1(E_s) + P_2^c \cdot \epsilon_2(E_s) + P_3^c \cdot \epsilon_3(E_s) \; ,
\end{eqnarray}
which we can write as a matrix equation:
\begin{equation}
 \vec{R} - \vec{P}_0 \cdot T_e(E_s) = \mathbf{P} \cdot \vec{\epsilon} \quad {\rm with} \quad
 \mathbf{P} = \left(\begin{array}{ccc} 
      P_1^a & P_2^a & P_3^a \\
      P_1^b & P_2^b & P_3^b \\
      P_1^c & P_2^c & P_3^c 
     \end{array}\right) \; .
\end{equation}
Taking higher scattering orders into account would require additional measurements at further non-zero column densities and would increase the dimension of the system of linear equations to be solved. Whether the inclusion of only three scattering orders provides sufficiently accurate results will be evaluated in section~\ref{sec:eval}.
\par
Multiplying with the inverse of $\mathbf{P}$, which is calculated using the Gauss-Jordan algorithm from the ROOT software package~\cite{root}, we obtain 
\begin{equation}
 \vec{\epsilon} = \mathbf{P^{-1}} \cdot \left(\vec{R} - \vec{P}_0 \cdot T_e(E_s)\right)
\end{equation}
from which we can calculate the single scattering function $\epsilon_1(E_s)$.
\subsection{Deconvolution of the energy loss function}
\label{sec:decon}
As described in section~\ref{sec:resp}, the single scattering function is the result of the convolution of the experimental transmission function of the spectrometer with the energy loss function. This convolution is calculated taking the integral
\begin{equation}
 \epsilon_1 (E_s) = T_e(E_s) \otimes f(\Delta E) = \int^{E/2}_{0} T_e(E_s-\Delta E)f(\Delta E) \, \rm d\Delta E \; .
\end{equation}
In our case, where the values of the functions in question are only known at $N$ equally distributed discrete measurement points defined by the applied retardation voltage $U_i$, the integral is replaced by a sum
\begin{equation}
 \epsilon_1 (E-qU_i) = \sum^{N-1}_{j=0} T_e(E-qU_i-\Delta E_j)f(\Delta E_j) \; .
\end{equation}
The latter equation can be rewritten in $N \times N$ matrix form 
\begin{equation}
 \vec{\epsilon_1} = \mathbf{T_e} \cdot \vec{f} \; ,
 \label{eq:deconmatrix}
\end{equation}
where the $\mathbf{T_e}$ matrix is constructed from the discrete transmission function $T_e(E_{s,i} = E-qU_i)$ as\footnote{The zeroes in the right upper corner of $\mathbf{T_e}$ are caused by the transmission condition $E-qU_i-\Delta E_j \geq 0$.}
\begin{equation}
 \mathbf{T_e} = \left( \begin{array}{llllll} %{p{16mm}p{16mm}p{16mm}p{6mm}p{10mm}p{13mm}}
      T_e(E_{s,0})   & 0              &              &              & \cdots & 0            \\
      T_e(E_{s,1})   & T_e(E_{s,0})   & 0            &              & \cdots & 0            \\
      T_e(E_{s,2})   & T_e(E_{s,1})   & T_e(E_{s,0}) & 0            & \cdots & 0            \\
      \vdots         & \vdots         & \vdots       & \vdots       &        & \vdots       \\
      T_e(E_{s,N-1}) & T_e(E_{s,N-2}) & \cdots       &              &        & T_e(E_{s,0})
   \end{array} \right) \; .
\end{equation}
One could now try to solve equation~\ref{eq:deconmatrix} by multiplying with the inverse of the $\mathbf{T_e}$ matrix. The latter, however, is close to being singular and cannot easily be inverted numerically. We therefore have to apply more sophisticated methods to deconvolve the energy loss function from the matrix equation. In the following two methods are applied to the problem: the so-called Singular Value Decomposition (SVD)~\cite{numrec} and the iterative Stabilized Biconjugate Gradient method~\cite{Sle93}.
\subsubsection{Singular Value Decomposition}
\label{sec:svd}
The Singular Value Decomposition (SVD) is a method to deal with systems of linear equations given by a matrix equation $\mathbf{A} \cdot \vec{x} = \vec{b}$ that are either singular or numerically very close to singular and is able to provide useful, although not necessarily unambiguous, solutions to the given problem. 
It is based on the theorem that any $M\times N$ matrix $\mathbf{A}$ with $M\geq N$ can be written as a product of an $M\times N$ column-orthogonal matrix $\mathbf{U}$, an $N\times N$ diagonal matrix $\mathbf{W}$ whose elements are the so-called singular values $w_i \geq 0$, and the transpose of an $N\times N$ orthogonal matrix $\mathbf{V}$~\cite{numrec}:
\begin{equation}
 \mathbf{A} = \mathbf{U} \cdot \mathbf{W} \cdot \mathbf{V^T} = 
 \mathbf{U} \cdot \left( \begin{array}{cccc} w_1 &&& 0\\ & w_2 \\ && \dots \\ 0 &&& w_N \end{array} \right) \cdot \mathbf{V^T} ~ .
 \label{eq:svd1}
\end{equation}
As $\mathbf{U}$ and $\mathbf{V}$ are orthogonal the inverse of equation~\ref{eq:svd1} can be written as
\begin{equation}
 \mathbf{A}^{-1} = \mathbf{V} \cdot \mathbf{W}^{-1} \cdot \mathbf{U^T} = \mathbf{V} \cdot [diag(1/w_i)] \cdot \mathbf{U^T} ~ .
\end{equation}
Problems arise when some of the singular values $w_i$ are either zero or so small that their values are dominated by numerical rounding errors. Using the SVD method it is still possible to construct an approximate solution vector $\vec{x}$ that will minimize the residual $r$ given by 
\begin{equation}
 r \equiv |\mathbf{A}\cdot \vec{x} - \vec{b}| ~.
\end{equation}
For that purpose all diagonal elements of $\mathbf{W}^{-1}$ where the singular values $w_i$ are below a chosen threshold value $w_{\rm thr}$ are set to zero, thereby removing infinite or problematically large matrix elements. The matrix constructed using the modified 
diagonal matrix $\mathbf{\tilde{W}}^{-1}$ is the so-called pseudoinverse matrix $\mathbf{\tilde{A}}^{-1}$, and the solution vector $\vec{x}$ is then given by
\begin{equation}
 \vec{x} \approx \mathbf{\tilde{A}}^{-1} \cdot \vec{b} = \mathbf{V} \cdot \mathbf{\tilde{W}}^{-1} \cdot \mathbf{U^T} \cdot \vec{b} ~ ,
\end{equation}
which, translated to our original problem of deconvoluting the energy loss function from the measured single scattering function (see equation~\ref{eq:deconmatrix}), becomes
\begin{equation}
 \vec{f} \approx \mathbf{\tilde{T}_e}^{-1} \cdot \vec{\epsilon_1} = \mathbf{V} \cdot \mathbf{\tilde{W}}^{-1} \cdot \mathbf{U^T} \cdot \vec{\epsilon_1} ~ .
\end{equation}
What remains to be settled is the optimal threshold value $w_{\rm thres}$ for suppression of the problematic singular values. 
This can only be determined by investigating the influence of the deconvolved energy loss function on the extracted neutrino mass values in simulated neutrino mass runs of the KATRIN experiment. Such a study, applying a toy Monte Carlo simulation of the experiment, is presented in section~\ref{sec:MC}.
\subsubsection{Stabilized Biconjugate Gradient method}
\label{sec:iter}
As an alternative to the SVD method we tested the so called Stabilized Biconjugate Gradient (Bi-CGSTAB) method described by Sleijpen and Fokkema~\cite{Sle93}. 
The bi-conjugate gradient method iteratively solves linear sets of equations $\mathbf{A} \cdot \vec{x} = \vec{b}$ where $\mathbf{A}$ is an $N\times N$ matrix. In each iteration step, labeled $k$, the approximate solution $\vec{x}_k$ is modified by some search correction that depends on the true residual $\vec{r}_k = \vec{b} - \mathbf{A} \cdot \vec{x}_k$ and some ``shadow residual'' $\vec{\tilde{r}_k}$
calculated using the transpose $\mathbf{A}^T$.
The residuals are forced to converge by making $\vec{r}_k$ orthogonal to the shadow residuals $\vec{\tilde{r}_j}$ for $j < k$.\\
For an in-depth description of the algorithm we refer to reference~\cite{Sle93}. In our simulations we use the implementation of the Bi-CGSTAB algorithm provided as part of the {\em Meep} software package for finite-difference time-domain simulations developed at the Massachusetts Institute of Technology (MIT)~\cite{meep}. As for the SVD method, the energy loss function resulting from a deconvolution using the Bi-CGSTAB algorithm is evaluated in a toy Monte Carlo simulation of the experiment, as presented in section~\ref{sec:MC}.
\section{Evaluation using toy Monte Carlo simulation}
\label{sec:MC}
\begin{figure}[h]
\centering
\includegraphics[width=0.48\textwidth]{transmission.pdf}
\includegraphics[width=0.50\textwidth]{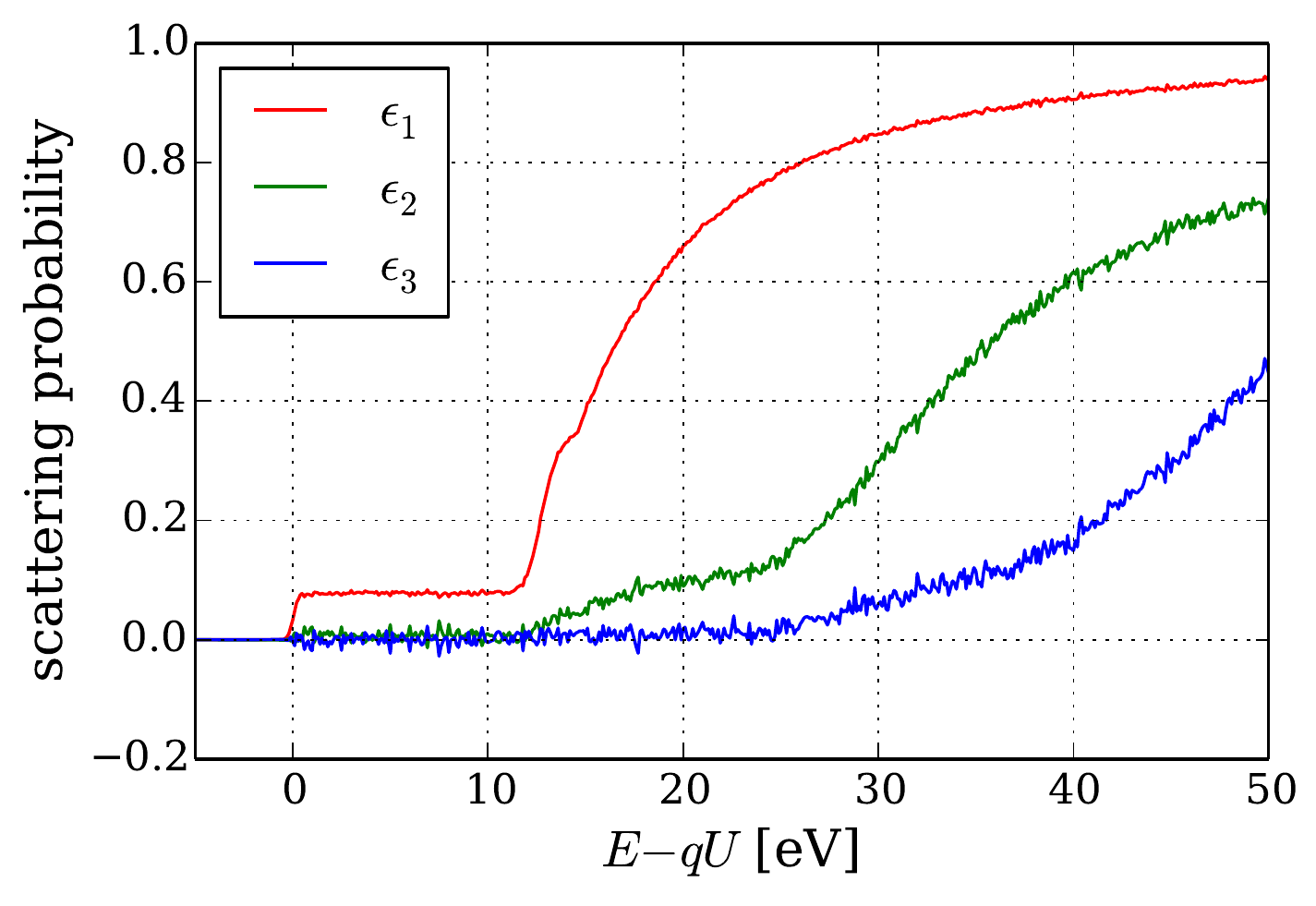}
\caption{Left: simulated response functions. Right: single, double and triple scattering functions extracted from the simulated response functions. The plateau of the red curve $\varepsilon_1$ is caused by the fraction of elastic scatterings. }
\label{fig:sim_res}
\end{figure}

In order to test the deconvolution methods described in section~\ref{sec:decon}, we performed Monte Carlo simulations of response function measurements at different column densities. In these simulations we assumed a perfectly homogeneous gas distribution within the WGTS and used the model of F.~Gl\"uck~\cite{Glueck} to generate energy losses in scattering events of 18.6~keV electrons with molecular hydrogen. Table~\ref{tab:simpar} provides an overview of the input parameters of the simulation.
\begin{table}[h]
 \centering
  \caption{Input parameters of the toy Monte Carlo simulation.}
 \begin{tabular}{lll}
  \toprule
  Parameter & Symbol & Value \\
  \midrule
  WGTS column densities          & $\rho_i d$             & $\{0, ~1 , ~3, ~5\} \cdot 10^{17}~{\rm cm}^{-2}$ \\
  Total inelastic cross section  & $\sigma_{\rm tot}$     & $3.7\cdot 10^{-18}~{\rm cm}^2$ \\
  Electron source energy         & $E$                    & 18.6~keV \\
  Electron source energy spread  & $\sigma_e$             & 0.2~eV \\
  Electron source maximum angle  & $\theta_{e,{\rm max}}$ & $0.5^\circ$ \\
  Electron source magnetic field & $B_e$                  & $3.6\cdot 10^{-2}$~T \\
  WGTS magnetic field            & $B_{\rm WGTS}$         & 3.6~T \\
  Analyzing plane magnetic field & $B_A$                  & $3.0\cdot 10^{-4}$~T \\
  Retardation voltage range      & $U_i$                  & 18550~V $\dots$ 18605~V \\
  Voltage step size               & $\Delta U_i$           & 0.1~V \\
  Number of electrons per step   & $N_e \pm \sqrt{N_e}$   & $1\cdot 10^7 \pm 3.16\cdot 10^3$ \\
  \bottomrule
 \end{tabular}
 \label{tab:simpar}
\end{table}
The simulated response functions are displayed in figure~\ref{fig:sim_res} (left). From these response functions we can 
extract the n-fold scattering functions as described in section~\ref{sec:scat} with the results shown in figure~\ref{fig:sim_res} (right). The single scattering function $\epsilon_1(E_s)$ is then the input for either one of the two deconvolution methods described in section~\ref{sec:decon}.
\subsection{SVD results}
\begin{figure}[h]
\centering
\includegraphics[width=0.48\textwidth]{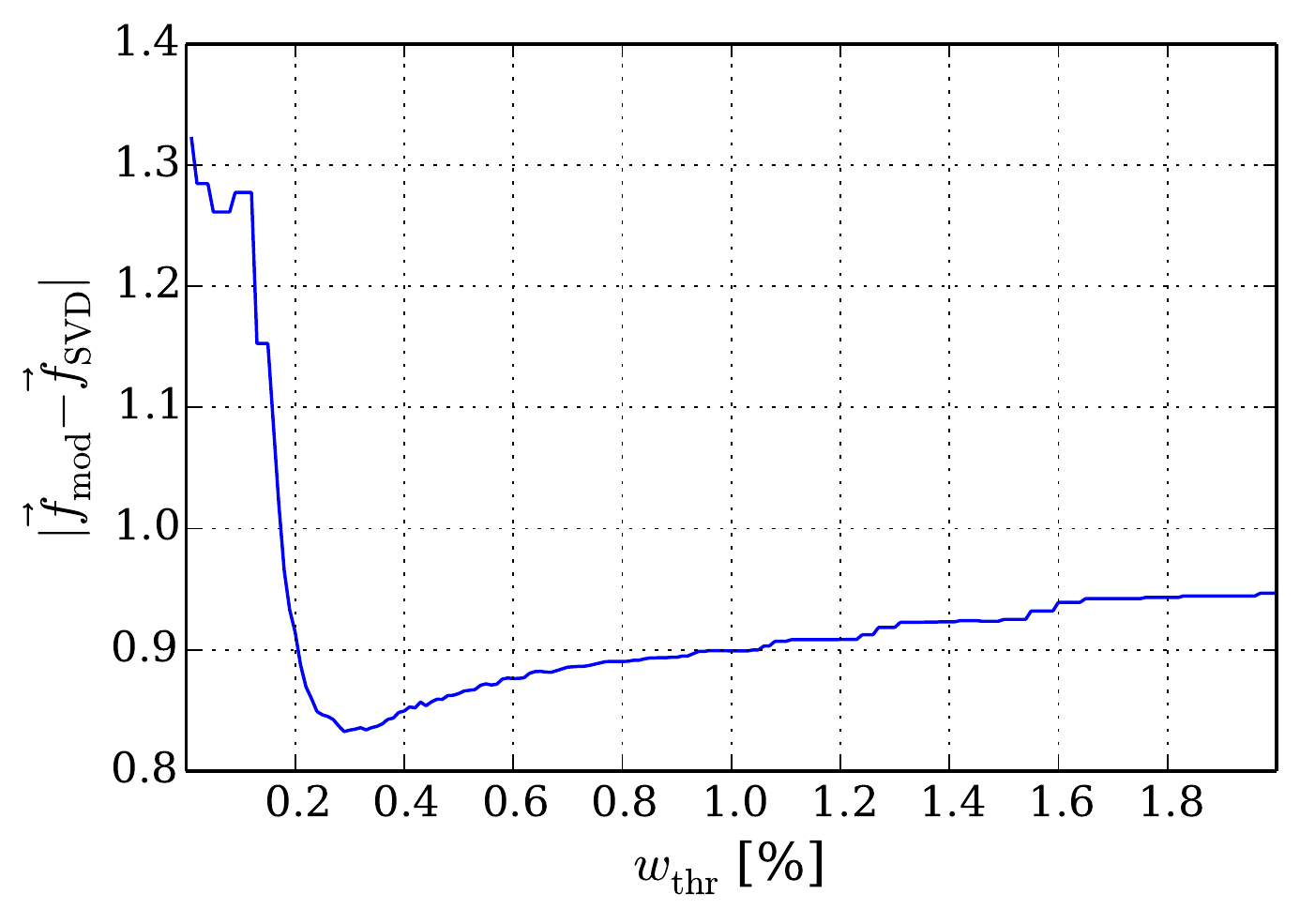}
\includegraphics[width=0.49\textwidth]{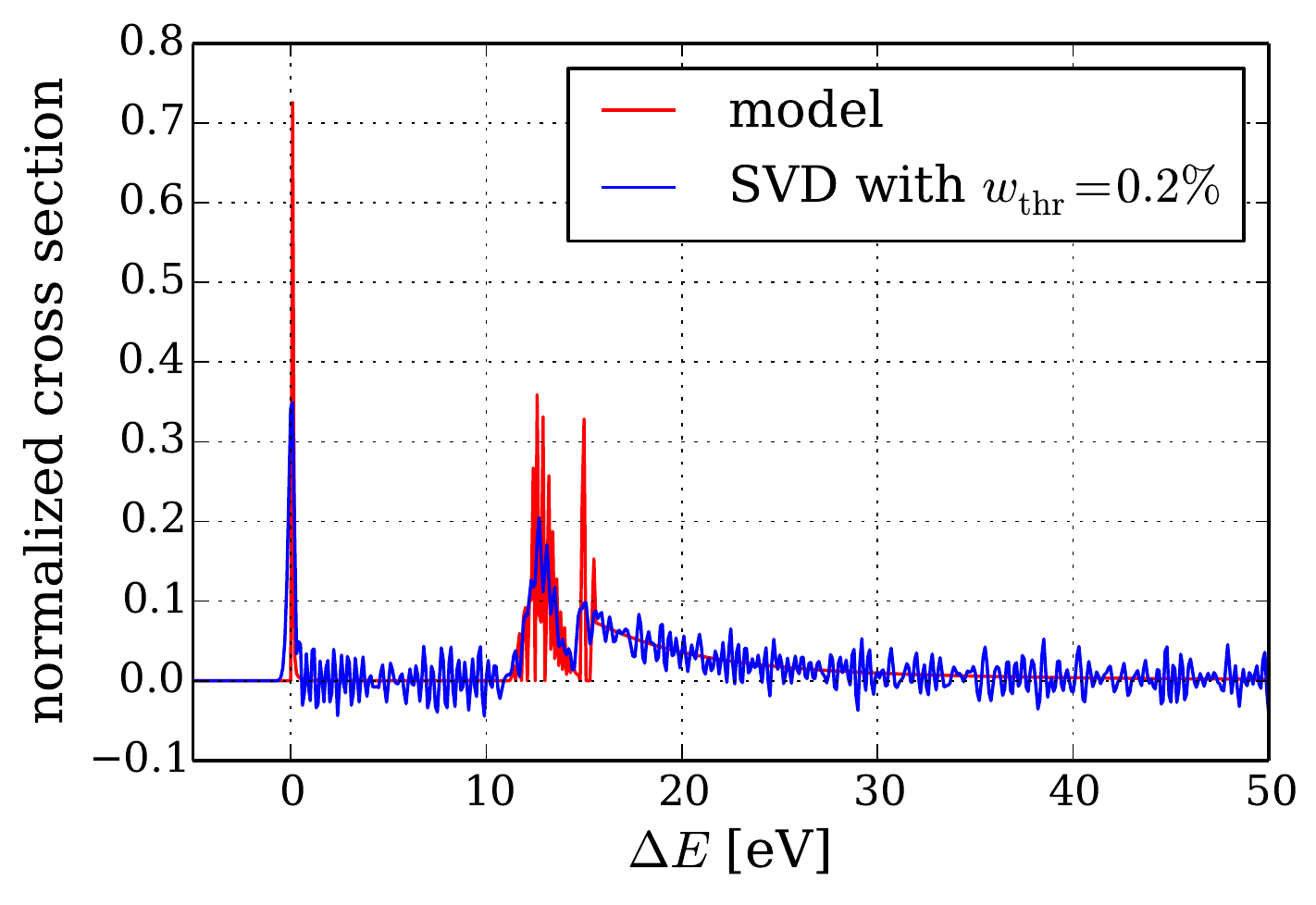}
\includegraphics[width=0.49\textwidth]{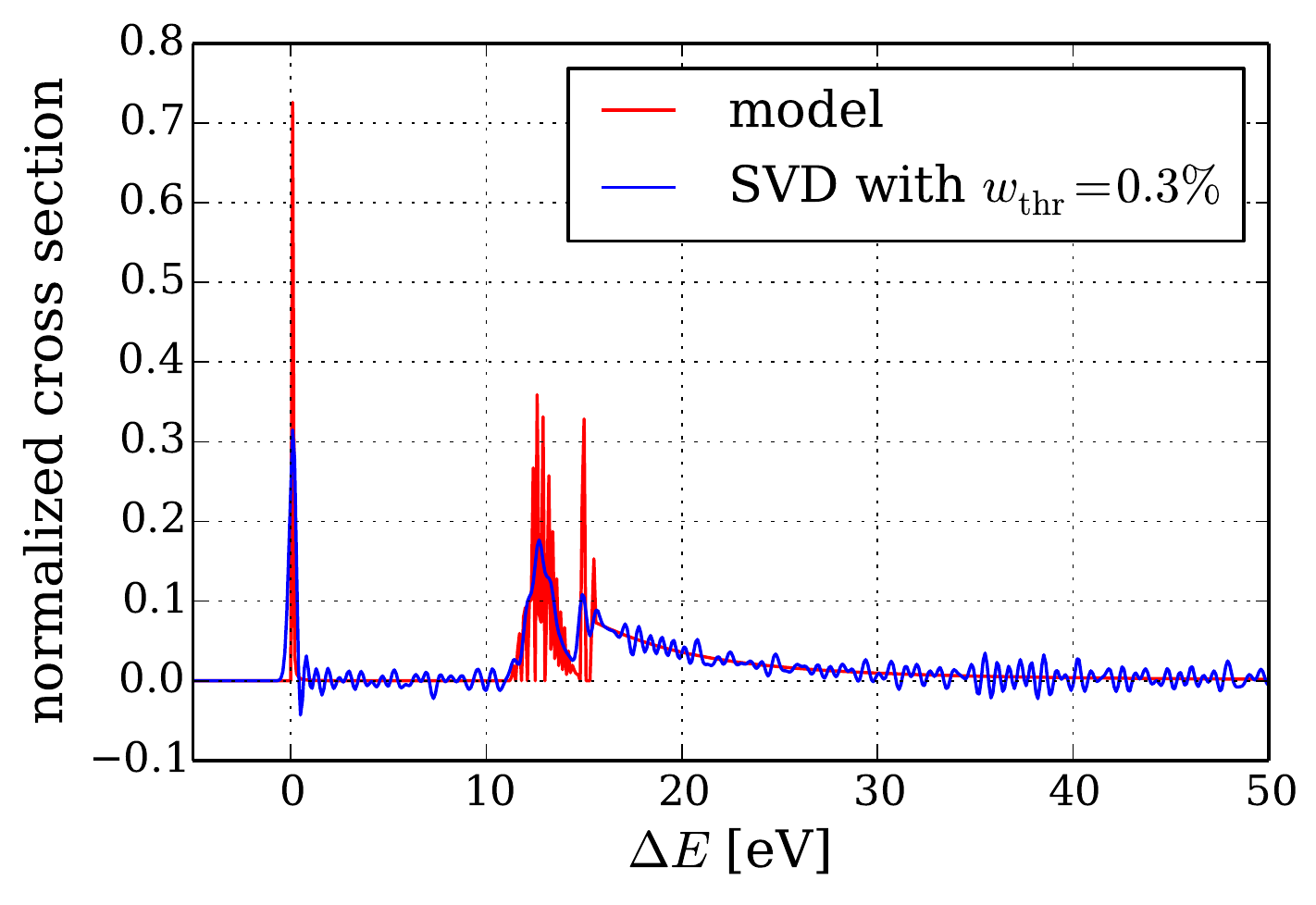}
\includegraphics[width=0.49\textwidth]{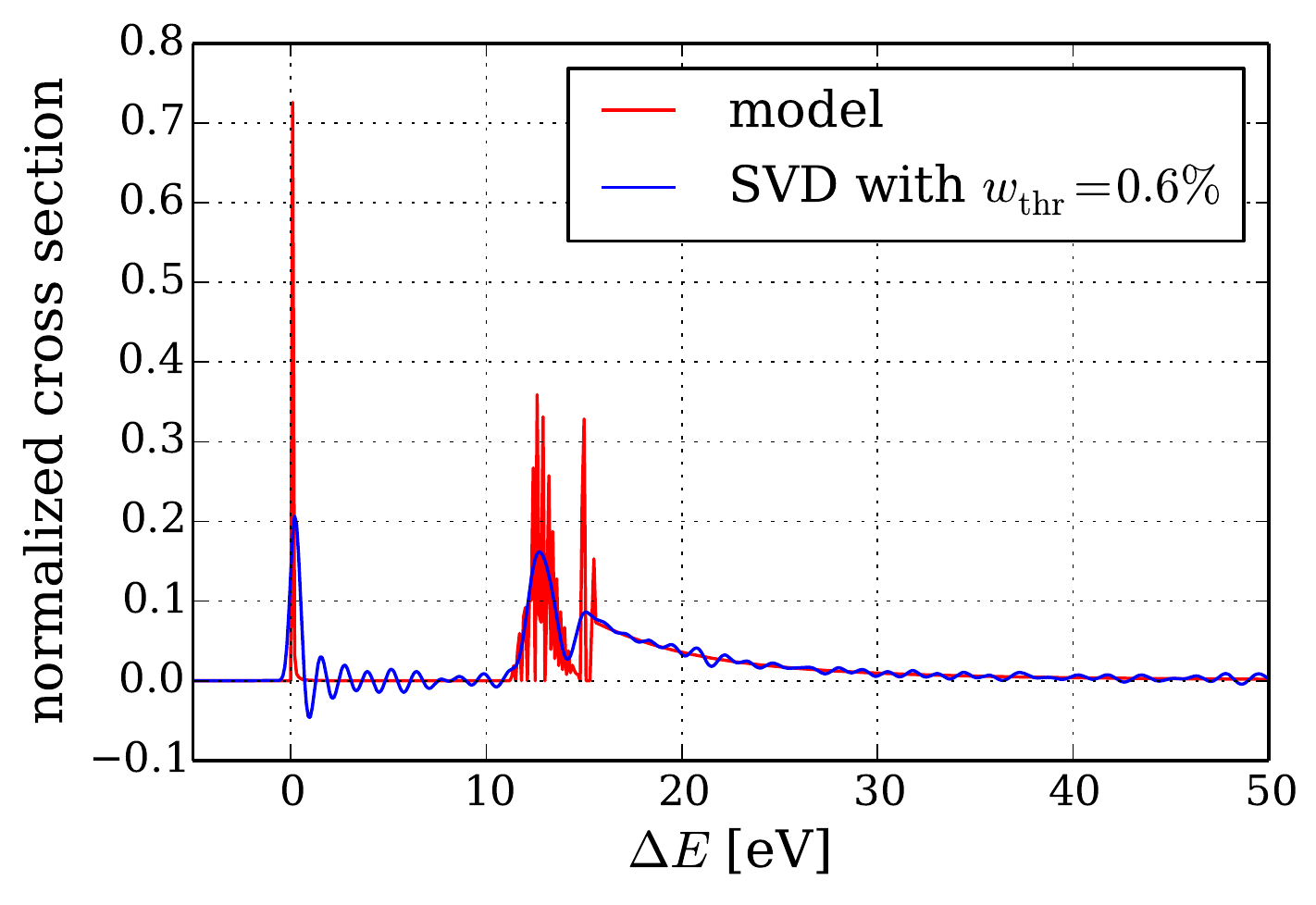}
\caption{Upper left: Determination of the optium threshold value. Other panels: Results of the SVD deconvolution of the energy loss function at different $w_{\rm thr}$ values. Shown in red is the input model of the energy loss function and in blue the deconvolved functions.}
\label{fig:svd_res}
\end{figure}

To optimize the result obtained for the energy loss function by the SVD method, we have to determine the optimum value of the threshold $w_{\rm thr}$ below which the corresponding matrix elements of the inverse diagonal matrix $\mathbf{W}^{-1}$ are discarded.
For that purpose we scan over a range of possible values for $w_{\rm thr}$ and with each value calculate the deconvolved energy loss function $\vec{f}_{\rm SVD}$ and the difference $|\vec{f}_{\rm mod} - \vec{f}_{\rm SVD}|$ to the input energy loss model $\vec{f}_{\rm mod}$. 
Figure~\ref{fig:svd_res} displays on the left the result of this scan. 
The threshold values are given in percent of the maximum singular value present in the diagonal matrix $\mathbf{W}$. 
Deconvolved energy loss functions obtained at threshold values of $0.2\%$, $0.3\%$ and $0.6\%$ are shown together with the input model. The suppression of low singular values in the SVD acts by damping numerical fluctuations in the deconvolved energy loss function. While the result obtained with $w_{\rm thr} = 0.2\%$ exhibits a significant amount of high frequency noise, a value of $w_{\rm thr} = 0.6\%$ produces a smoother result at the expense of washing out structures observed in the input model.
Besides this smoothening of the deconvolved function we also note an oscillatory response to large spikes in the input model, that is best visible in the lower right plot of figure~\ref{fig:svd_res}.
An optimum resemblence of the deconvolved function to the input model is found at a threshold value of $w_{\rm thr} = 0.3\%$.

It should be noted that the fact, that the SVD method cannot yield the fine structure of the energy loss function (see figure \ref{fig:eloss_fcn}) in the interval $11 \mathrm{eV} \leq \Delta E \leq 16 \mathrm{eV}$ at that threshold value, does not matter, since KATRIN`s transmission function for an isotropic electron source as the tritium source will have a width of 0.93~eV. We will investigate the influence on the neutrino mass measurement in section 4.3.
\subsection{Bi-CGSTAB results}
\begin{figure}[h]
\centering
\includegraphics[width=0.7\textwidth]{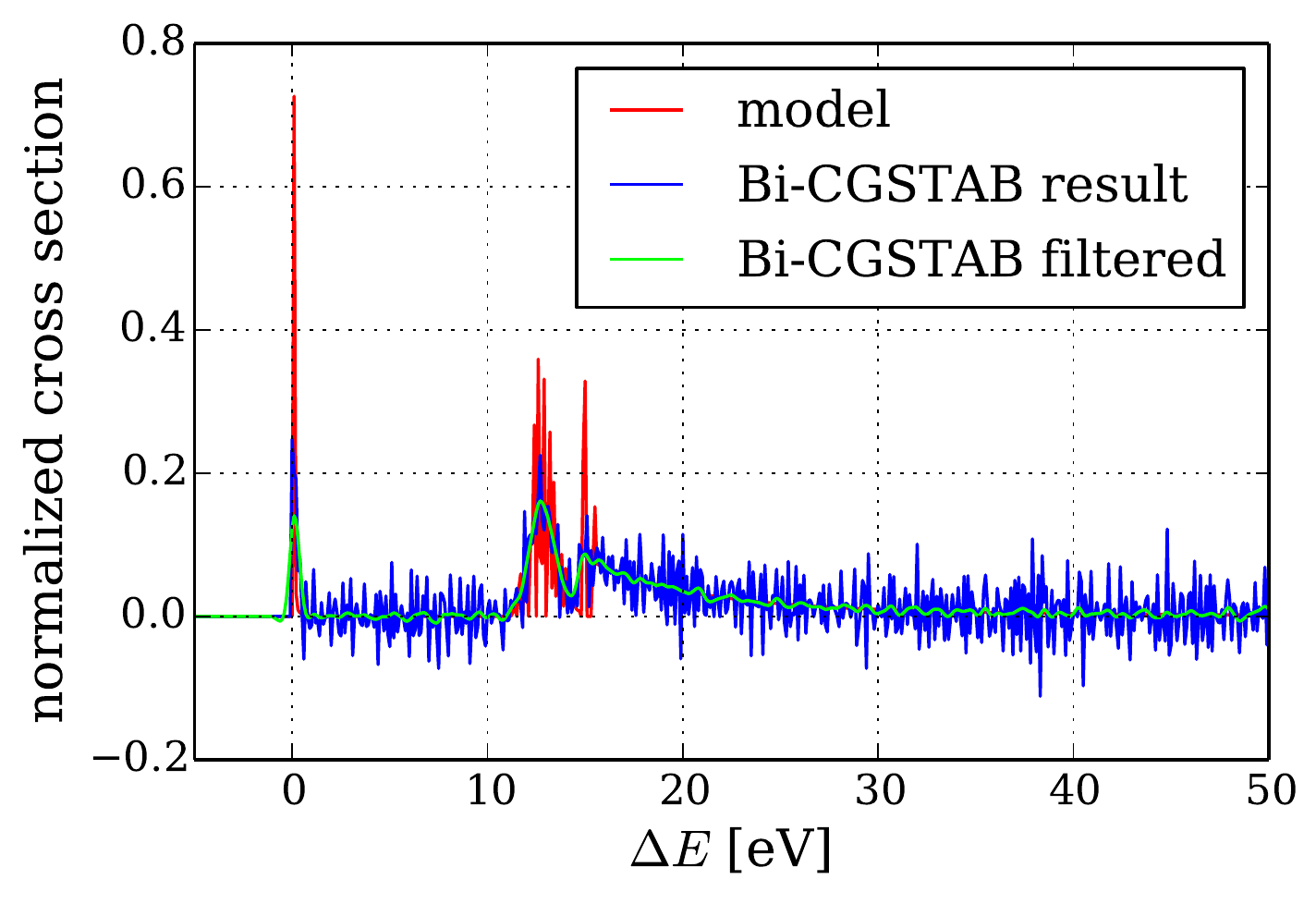}
\caption{Result of the iterative deconvolution of the energy loss function using the Bi-CGSTAB algorithm. Shown in red is the input model of the energy loss function and in blue the deconvolved function. A filtered version of the deconvolved function is overlaid in green.}
\label{fig:iter_res}
\end{figure}

Figure~\ref{fig:iter_res} displays the result obtained for the energy loss function using the iterative Bi-CGSTAB algorithm described in section~\ref{sec:iter}.
While the method does seem to reproduce some of the fine-grained structure related to inelastic electron excitations starting at $\Delta E \geq 12$~eV, it is obviously much noisier than the results obtained with the SVD method. To counteract the noise, we tried to filter the deconvolved function with a second-order low pass Butterworth filter with a cut off frequency of 1~eV$^{-1}$, with the result shown in green in figure~\ref{fig:iter_res}. The filtered function does exhibit lower noise and still follows the general features of the input model. 
\subsection{Influence on the measured neutrino mass}
\label{sec:eval}
\begin{figure}[h]
\centering
\includegraphics[width=0.49\textwidth]{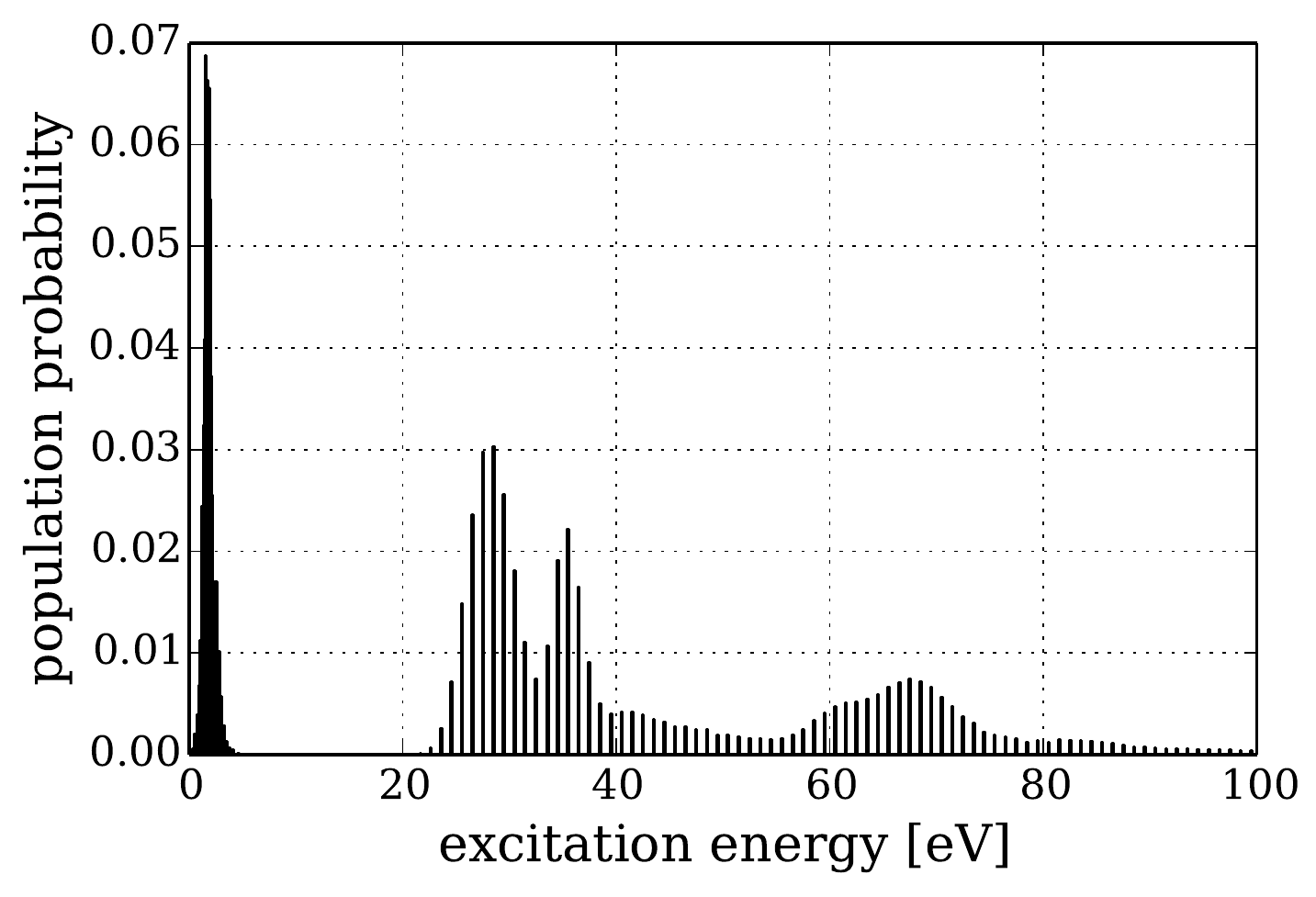}
\includegraphics[width=0.49\textwidth]{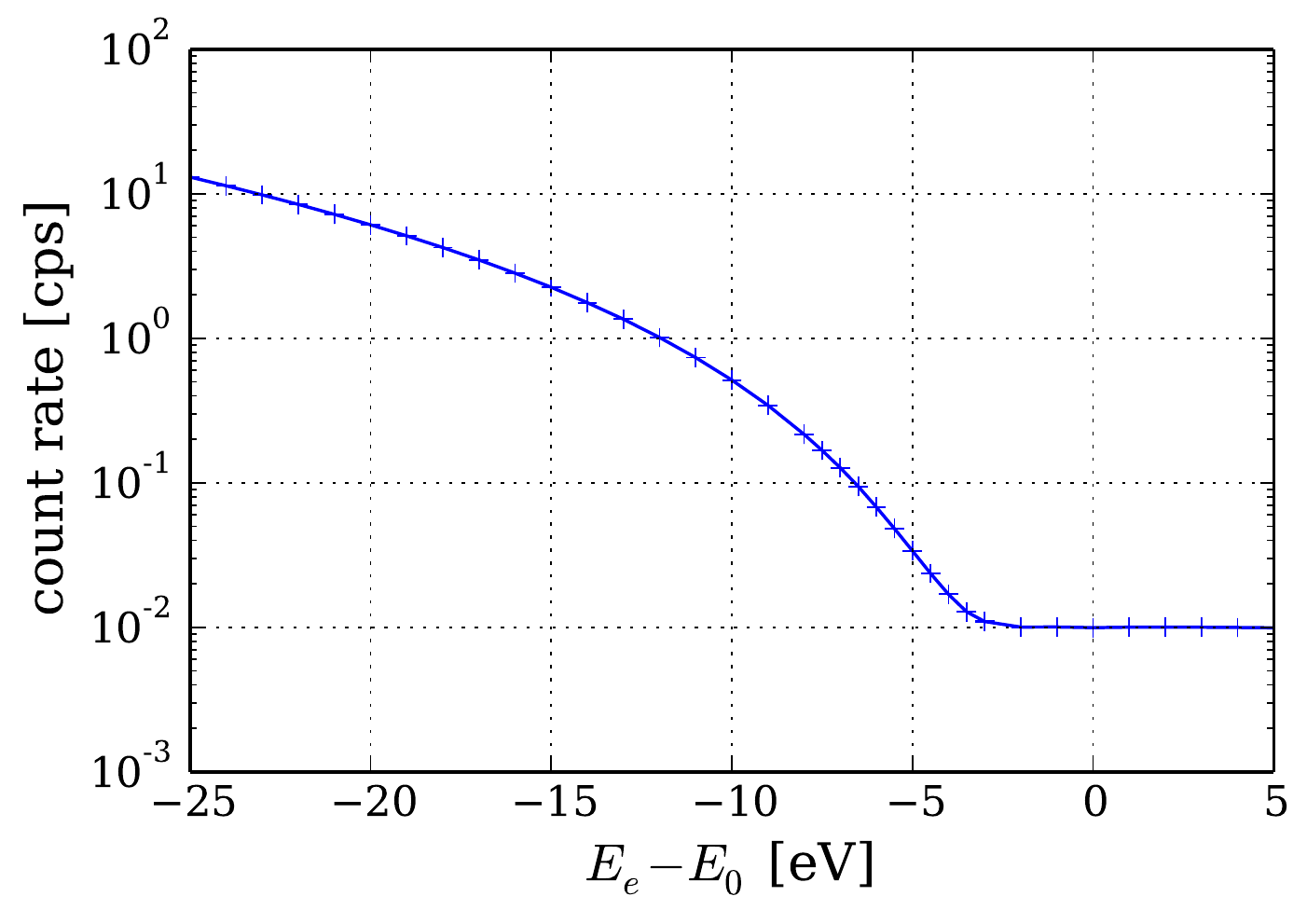}
\caption{Left: Assumed final state distribution of the $^3$HeT$^+$ daughter molecules in the $\upbeta$ decay (adopted from \cite{Sae00}). Right: simulated integral $\upbeta$ spectrum.}
\label{fig:betaspectrum}
\end{figure}

To assess the influence of the deconvolution process on the determination of the neutrino mass from the experimental beta spectrum measured by the KATRIN experiment, a large number of such measurements have been simulated and analyzed using the different energy loss functions displayed in figures~\ref{fig:svd_res} and~\ref{fig:iter_res}. From the mean of the distribution of the extracted $\mnu^2$ values we can then estimate the systematic uncertainty caused by the deconvolution process.\\
The energy spectrum of nuclear $\upbeta$ decay can be calculated starting from Fermi's golden rule and has the following form~\cite{Ott08} (with units $\hbar = c = 1$):
\begin{eqnarray}
 \frac{\rm d\Gamma}{\rm dE}  & = & \frac{G_{\rm F}^2 \cos^2\theta_{\rm C}}{2\pi^3}  |M|^2 \, F(E,Z+1) \, p \, (E + m_e)  \nonumber \\
 && \cdot \sum_j \mathcal{P}_j \; E_{\upnu,j} \; \sqrt{E_{\upnu,j}^2-\mnu^2} \; \Theta(E_{\upnu,j} - \mnu) \; ,
\label{eq:betaspectrum}
\end{eqnarray}
where $G_{\rm F}$ is the Fermi coupling constant, $\theta_{\rm C}$ the Cabibbo angle and $M$ the nuclear matrix element of the transition. The Fermi function $F(E, Z+1)$ takes into account the final state interaction of the emitted electron with the daughter nucleus of charge $Z+1$ and $p(E+m_e)$ is the phase space factor of the outgoing electron. The product of the neutrino momentum and its energy given by $E_{\nu,j} = E_0-E^*_j- E$ is the phase space of the emitted anti-neutrino, which shapes the $\upbeta$ spectrum near its endpoint. The neutrino phase space factor has to be summed up over all final states $E^*_j$ of the daughter molecule that are populated with probabilities $\mathcal{P}_j$. The inclusion of the $\Theta$ function in eq.~\ref{eq:betaspectrum} ensures that $E_{\upnu,j} - \mnu > 0$.
The observable $\mnu^2$ that can be extracted from the spectral shape near the endpoint is defined by an incoherent sum over the neutrino mass eigenstates $m_i$ weighted by the matrix elements of the $U_{\rm{PMNS}}$ mixing matrix~\cite{PDG14} known from oscillation experiments:
\begin{equation}
 \mnu^2 = \sum^3_{i=1} | U_{ei} |^2 m_i^2 \; .
\label{eq:numass}
\end{equation}
The simulation of the observed integral spectrum measured by KATRIN uses the electron spectrum described by equation~\ref{eq:betaspectrum} and takes into account a number of experimental effects:
\begin{itemize}
 %\item A zero neutrino mass. I'd prefer not to list this here, as it is not an experimental effect. See blow.
 \item The final state distribution of the $^3$HeT$^+$ daughter molecules calculated by Saenz et al.~\cite{Sae00} as shown in figure~\ref{fig:betaspectrum} (left).
 \item Up to four-fold scattering of the electrons within the WGTS according to the same energy loss model as used in the simulations of the deconvolution process.
 \item The nominal transmission function of the main spectrometer.
 \item An experimental background rate of 10$^{-2}$ counts per second in the energy region of interest at the FPD.
 \item Statistical fluctuations according to an effective three years worth of data taken at the design value for the column density of the source of $5\times 10^{17}/{\rm cm}^2$ and with a measurement time distribution as described in the KATRIN Design Report~\cite{KAT04}. 
\end{itemize}
For the purpose of this study, a vanishing neutrino mass $\mnu = 0$ is assumed.
With these inputs, 1000 hypothetical KATRIN measurements were simulated and subsequently fitted taking into account the same physical effects as in the simulations, but using one of the deconvolved functions to describe the energy losses and using as free parameters 
\begin{itemize}
\item the squared neutrino mass $\mnu^2$,
\item the spectral endpoint energy $E_0$,
\item the experimental background rate, and
\item the overall amplitude of the spectrum.
\end{itemize}
    
\begin{figure}[h]
 \centering
 \includegraphics[width=0.49\textwidth]{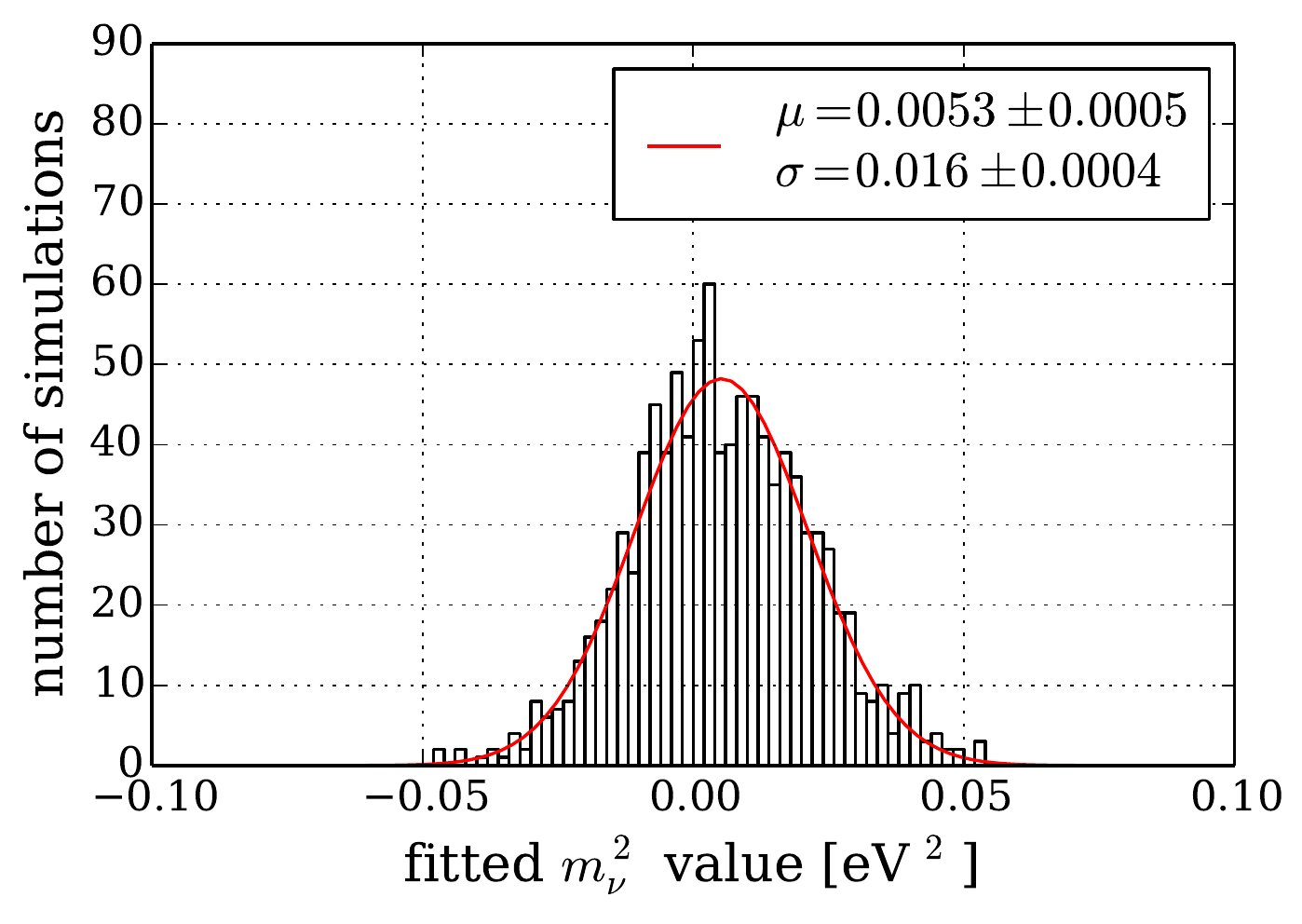}
 \includegraphics[width=0.49\textwidth]{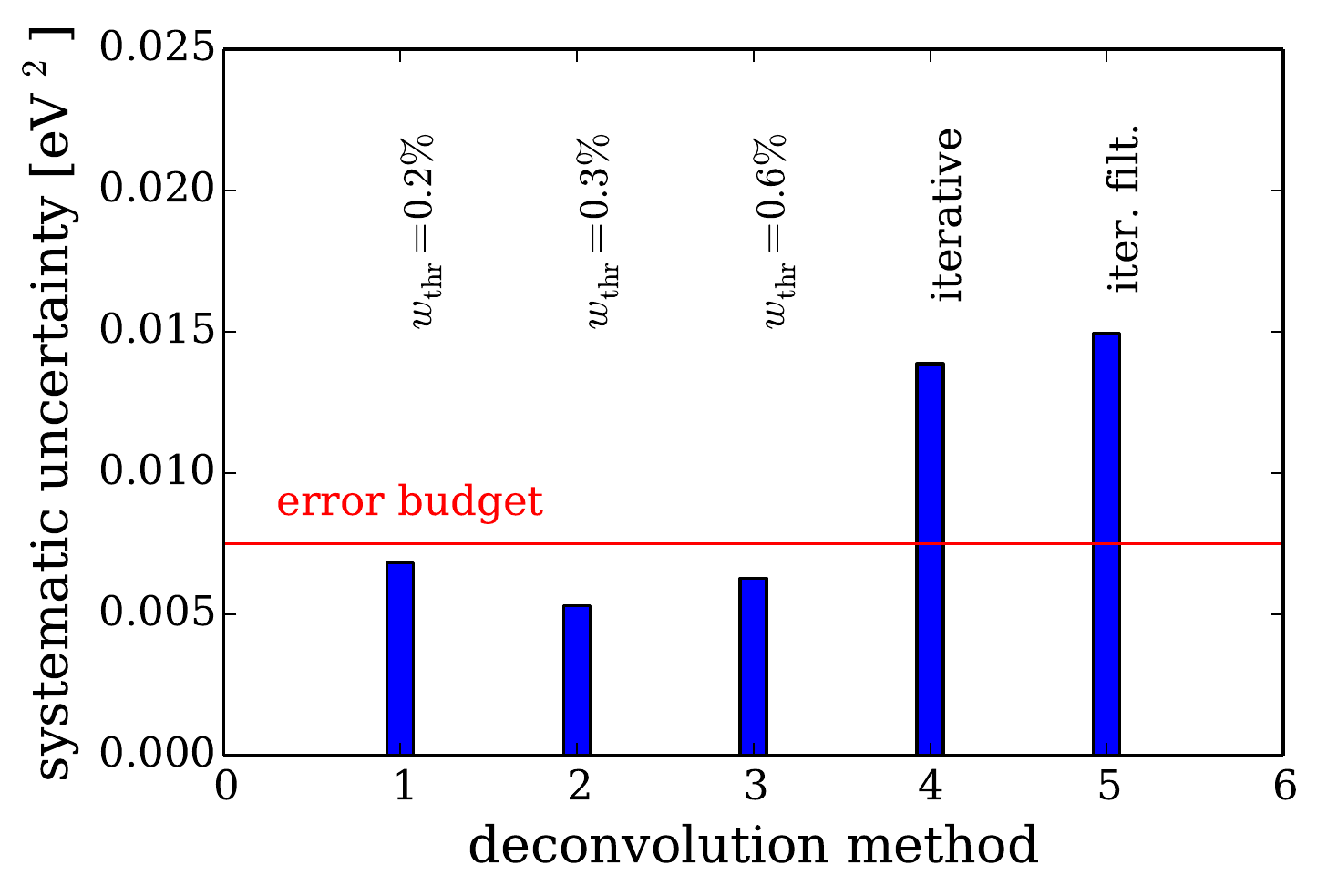}
 \caption{Left: distribution of $\mnu^2$ values extracted from the fits of simulated KATRIN spectra using the SVD method with $w_{\rm thr} = 0.3\,\%$. Right: systematic errors induced by the use of energy loss functions deconvolved with different SVD threshold values and with the iterative Bi-CGSTAB algorithm.}
 \label{fig:ind_err}
\end{figure}

Figure~\ref{fig:ind_err} (left) shows the resulting distribution of extracted $\mnu^2$ values obtained with the 
energy loss function from the SVD deconvolution at a threshold value of $w_{\rm thr} = 0.3~\%$. The mean of the distribution is at $\mu = (0.0053 \pm 0.0005)~{\rm eV}^2$ yielding the systematic uncertainty of the deconvolution method, compatible with the KATRIN error budget for this systematic effect of $0.0075~{\rm eV}^2$~\cite{KAT04}. The systematic uncertainties corresponding to usage of the energy loss functions obtained with the different threshold values of the SVD algorithm and with the iterative Bi-CGSTAB algorithm (with and without filter), respectively, are compared in figure~\ref{fig:ind_err} (right).
The energy loss functions deconvolved using the SVD algorithm result in significantly lower systematic errors than with the iterative Bi-CGSTAB algorithm and are, for all three threshold values tested, within the error budget defined for this uncertainty in the KATRIN design report~\cite{KAT04}.
As a cross check, the fits were also run using the input model itself for the energy loss function and the resulting $\mnu^2$ distribution was indeed centered around zero with $\mu = (0.0003 \pm 0.0005)~{\rm eV}^2$.
\section{Summary and Outlook}
A method to deconvolve the energy loss function of $\upbeta$-decay electrons in the tritium source of the KATRIN experiment from measurements of the response function at different column densities with a mono-energetic electron source has been developed and tested in simplified Monte Carlo simulations of the experiment. Two different algorithms to deconvolve the energy loss function from the single scattering function extracted from the measurements have been tested: Firstly, the so-called Singular Value Decomposition (SVD), and secondly, the Stabilized Biconjugate Gradient (Bi-CGSTAB) method as an example of an iterative algorithm.
Both methods allow to obtain an approximation of the real energy loss function with varying levels of detail and of numerical fluctuations in the deconvolved result. The numerical noise can be dampened in the case of the SVD algorithm by setting an appropriate threshold value for the suppression of small singular values and in the case of the Bi-CGSTAB algorithm by filtering the deconvolved function. When applying the deconvolved energy loss function to the analysis of simulated integral $\upbeta$ spectra, we find that the SVD method delivers usable results that induce a systematic error below the error budget of $0.0075~{\rm eV}^2$ allocated for this contribution in the KATRIN design report. The Bi-CGSTAB method, however, significantly exceeds the limit, both for the direct and for the filtered result of the deconvolution.
\par
An option to improve the result from SVD deconvolution that has been looked at during simulations is to split the energy loss function into several energy intervals and perform a sectionwise deconvolution of the data. In this way it is possible to choose lower threshold values in the regions of rotational and vibrational states and of inelastic electron excitations where more structure is expected and to dampen numerical fluctuations by higher threshold values in between these regions and in the ionization tail, where we know that the curve is smooth.
With such an approach it was indeed possible to get a closer approximation of the fine structure observed in the input model of the energy loss function. However, the energy loss function deconvolved in this manner led to larger systematic errors in the fits of the neutrino mass spectra and was therefore discarded.
\par
Further improvements of the deconvolved results are possible by increasing the statistics of the response function measurements. 
In the simulations we have assumed to collect 10 million events at each voltage step of the measurements. Assuming an event rate of 25~kHz this would result in about 10 days measurement time plus overhead, e.g., for switching between voltage settings during the measurements and for changing the column density of the WGTS in between runs. 
Moreover, one might consider including four-fold scattering in the extraction of the single scattering function (see section~\ref{sec:scat}) to increase the precision of the subsequently deconvolved energy loss function. However, also this action comes at the cost of increased measurement time, as this would require to measure at yet an additional, higher column density setting of the source.
Finally it could be investigated if going beyond the present precision might be possible by applying Bayesian methods as used to reconstruct spectral functions in lattice QCD~\cite{burnier13}.
\section*{Acknowledgments} 
This work was supported by the German Federal Ministry of Education and Research (BMBF) under grant no. 05A11PM2 and 05A14PMA. 
KV wishes to acknowledge support by the Helmholtz Association (Young Investigators Group VH-NG-1055).
\section*{References}
{}
\end{document}